\documentclass[twocolumn]{jpsj2}
\title{Skutterudite Results Shed Light on Heavy Fermion Physics}

\author{%
E.~\textsc{Hassinger}$^1$,
J.~\textsc{Derr}$^1$,
J.~\textsc{Levallois}$^2$,
D.~\textsc{Aoki}$^1$,
K.~\textsc{Behnia}$^3$,
F.~\textsc{Bourdarot}$^1$,
G.~\textsc{Knebel}$^1$,
C.~\textsc{Proust}$^2$ and
J.~\textsc{Flouquet}$^1$
}

\inst{%
$^1$DRFMC/SPSMS, Commissariat a l'Energie Atomique, F-38042 Grenoble, France\\
$^2$Laboratoire National des Champs Magnetiques Pulses (CNRS), BP 4245, 31432 Toulouse, France\\
$^3$Laboratoire de Physique Quantique (CNRS), ESPCI, 10 Rue de Vauquelin, 75231 Paris, France\\
}

\abst{%
Only few selected examples among the great diversity of anomalous rare earth skutterudite are reviewed. Focus is first given on PrFe$_4$P$_{12}$ in comparison with URu$_2$Si$_2$. For PrFe$_4$P$_{12}$, great progress has been made on determining the nature of the order parameter (OP). A non magnetic order parameter with a multipolar component emerges here while for URu$_2$Si$_2$ the nature of the so-called hidden order remains mysterious. The two systems have several similarities in their temperature--pressure ($T$, $P$) and magnetic field--temperature ($H$, $T$) phase diagrams, in their spin dynamics, in their nesting character and in their high sensitivity to impurities. Advances on one side must stimulate new views on the other. Besides general considerations on the choice of the OP, a simple basic problem is the treatment of the Kondo coupling in a system with low charge carrier number for the cases of uncompensated and compensated semi-metal. An interesting problem is also the possible decoupling between exciton modes and itinerant carriers.
}

\kword{}

\begin{document}
\maketitle

The new skutterudite compounds such as PrFe$_4$P$_{12}$, SmOs$_4$Sb$_{12}$, SmFe$_4$P$_{12}$, CeOs$_4$Sb$_{12}$ and CeFe$_4$P$_{12}$ give new insights on the interplay between valence driven metal--insulator transitions, Kondo effect and magnetic long range order. A fancy point is that these compounds often seem to have an intermediate valence behaviour even when their valences are close to 3. The occurrence of well defined multipoles is often revealed by crossing the phase transition in temperature. Magnetic field will lead to recover the duality between localized and intermediate valence characters. To illustrate this particularity, a comparison is made with electronic conduction and magnetism in the intermediate valence system SmB$_6$ and TmSe and also with the ferromagnetic superconductor URhGe.

\section{Peculiarity of skutterudites in the heavy fermion family}
The Ce, Pr or Sm base skutterudite systems add new milestones in the domain of heavy fermion compounds (HFC), notably on the occurrence of a multipolar order parameter (OP). One of the key ingredients is the tendency of the isostructural lanthanum host to nesting with a characteristic vector $(1, 0, 0)$. This particularity leads to the rare situation that a complex system can lead to singular new clear properties~\cite{YAok05}.

The achievement of heavy--fermion behaviour is due to the strong hybridization with neighbour atoms in these cage systems. It leads to a relatively high Kondo temperature ($T_{\rm K}$) 
even for an occupation number ($n_f$) of the trivalent configuration close to $n_f = 1$. Even for $n_f \sim 1$ a large effective mass $m^\ast$ of the quasiparticle occurs,  as if the rare earth substitution of La atoms boosts the formation of heavy quasiparticles.

The effect is reinforced by the weakness of the crystal field (CF) splitting $C_{\rm CF}$~\cite{Kur05} with $k_{\rm B}T_{\rm K} \sim C_{\rm CF}$. The system looks like an intermediate valence compound in the paramagnetic (PM) state with strongly damped CF excitations. For example, the inelastic neutron scattering spectrum shows a monotonous behaviour with no indication of defined excitations in frequency and wavevector (see results on Ce intermediate valence systems like CeSn$_3$ and CeBe$_{13}$ in refs.~\citen{Mur86,Loe93}).

However, suddenly, when a phase transition at $T_{\rm A}$  occurs the local $4f$ character is revealed in association to drastic changes of the Fermi surface. Often, on cooling below $T_{\rm A}$, a decoupling between exciton type excitations characteristic of the OP and of the bare crystal field scheme on one hand and surviving itinerant carriers  on the other hand appears. An excellent example for such a scenario is PrFe$_4$P$_{12}$. 

External tuning via pressure ($P$) or magnetic field ($H$) will produce the usual effects known for HFC (change of $T_{\rm K}$ related to the change of $n_f$, concomitant a change of $C_{\rm CF}/k_{\rm B}T_{\rm K}$ leading to a modification in the nature of the interaction). The novelty is that due to the weakness of $C_{\rm CF}$, drastic changes in the CF scheme can occur which generates a strong feedback on the nature of the ordered multipole; the different multipoles are now field dependent.
A key feedback is the possible variation of the carrier number $n$,
which can happen at the phase transition.

The aim of the present article is: (1) to stress the strong input of anomalous rare earth skutterudites in the study of HFC problems by selecting the recent example of PrFe$_4$P$_{12}$, (2) to compare this emerging almost resolved case to the ``mysterious'' situation of URu$_2$Si$_2$  where the problem of its so called hidden order phase remains greatly open, (3) to discuss the interplay between valence, electric conduction and multipolar order parameter (multipolarity, nature of the magnetic interactions). On this last point, the HFC points of  reference will be  Sm hexaboride and Tm calcogenides.  The skutterudite examples will be CeOs$_4$Sb$_{12}$, which is a Kondo insulator close to an antiferromagnetic instability, and CeFe$_4$P$_{12}$, an apparently robust PM insulator. SmOs$_4$Sb$_{12}$ and SmFe$_4$P$_{12}$ are two metallic Kondo systems close to ferromagnetic instability. It is an open issue whether they can become superconductors like UGe$_2$ or URhGe~\cite{Flo05}. The interesting case of exotic superconductivity (PrOs$_4$Sb$_{12}$) in strongly fluctuating multipolar medium is not discussed. Different points of view can be found in the proceedings.

\section{PrFe$_4$P$_{12}$: a reference with multipolar order parameter}
The beauty of PrFe$_4$P$_{12}$  is that recent microscopic measurements have clarified the symmetry of the OP with a continuous theoretical feedback. Below a temperature named as $T_{\rm A}$, PrFe$_4$P$_{12}$ is a low carrier semi-metal with a charge carrier number $n$ per Pr formula between $10^{-3}$--$10^{-2}$, while above $T_{\rm A}$ it is considered to be an excellent metal with a carrier number $n\sim 1$~\cite{YAok02,Pou06}. The observation of a structural modulation with wavevector $(1, 0, 0)$ indicates the loss of $(1/2, 1/2, 1/2)$ translation linked with no net sublattice magnetization~\cite{Iwa02}. NMR analysis~\cite{Kik07} of  totally symmetric magnetic multipoles induced by magnetic field clarifies that, below $T_{\rm A}$, the ordered phase A has broken translational symmetry which preserves the $T_h$ point symmetry of the Pr sites. All theoretical models converge with the scheme of a multipolar order parameter unfortunately referred by different names such as scalar order parameter~\cite{Kis06}, AF monopole~\cite{Sak07} or AF hexadecapole~\cite{Tak06}. We will use the label of nmOP (non magnetic order parameter) for the phase A waiting for a well defined label by the experts.

\begin{figure}[htb]
\begin{center}
\includegraphics[width=0.8 \hsize,clip]{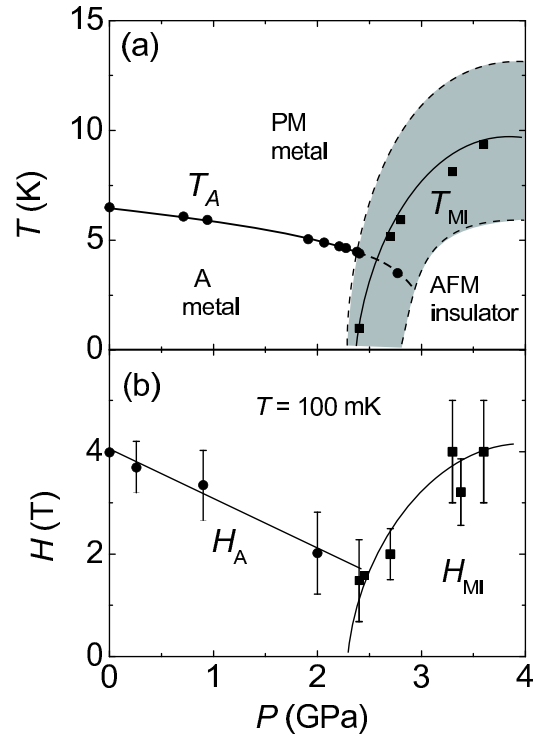}
\end{center}
\caption{
(a) $(T, P)$ phase diagram of PrFe$_4$P$_{12}$, cited from ref.~\protect\citen{Hid05}. A is the nmOP semimetallic phase, at high pressure the ground state switch to an AF insulator phase. The gray area indicates data obtained with different $P$ hydrostaticity;
(b) $(H, P)$ phase diagram of PrFe$_4$P$_{12}$  at $T \to 0\,{\rm K}$
}
\label{fig:1}
\end{figure}

Under pressure, the $(1, 0, 0)$ lattice modulation disappears. There is a symmetry lowering from cubic to at least orthorhombic~\cite{Kaw06} associated to a metal-insulator transition and an (AF) antiferromagnetic order with a propagation vector $(1, 0, 0)$~\cite{Osa07} as represented in Fig.~\ref{fig:1}(a)~\cite{Hid05}.

As the magnetic field response will reflect the different shifts of crystal field levels it is not surprising that whatever the ground state A or AF, the restoration of a good metal with a large number of carrier occurs at a quite comparable value $H_{\rm A}$ or $H_{\rm MI} \sim 4\,{\rm T}$ of the magnetic field. (Fig.~\ref{fig:1}b). It is also interesting to remark that the achievement of noticeable magnetic polarization (magnetization per Pr atom $\sim 1\,\mu_{\rm B}$ at $H_{\rm A}$) feeds to a metallic conduction. An interesting feature is that a relative weak substitution of Pr by La is sufficient to destroy the A phase and to lead to ferromagnetism (FM)~\cite{Tay07}.

Just above $T_{\rm A}$, PrFe$_4$P$_{12}$  appears at least for its spin dynamic to be an intermediate valence system with a structureless broad spectrum while below $T_{\rm A}$  nicely defined excitations develop~\cite{Par05,Koh06}, which must be characteristic of its nmOP. The beauty of the system is that the main part of the Fermi surface collapse at $T_{\rm A}$  (the a$_{\rm u}$ band) while the small a$_{\rm g}$ band survives~\cite{YAok05,Har03}. The quasiparticles which remain itinerant move in a neutronal medium formed by the excitons associated with the nmOP. Below a temperature $T_{\rm x} \sim 2.8\,{\rm K} < T_{\rm A}  \sim 6.5\,{\rm K}$~\cite{Pou06,Pou_proc}, itinerant quasiparticles and the excitons seem to form two different entities as $^3$He in superfluid $^4$He. Qualitatively $T_{\rm x}$ is the result of the lowering of the Fermi temperature $T_{\rm F}$ with the strong decrease of the electronic carrier number $T_{\rm F} \sim n^{2/3}$ and of course of the dressing (effective mass) by their motion through the solid.

\section{URu$_2$Si$_2$: still the mystery of its OP}
\begin{figure}[htb]
\begin{center}
\includegraphics[width=0.8 \hsize,clip]{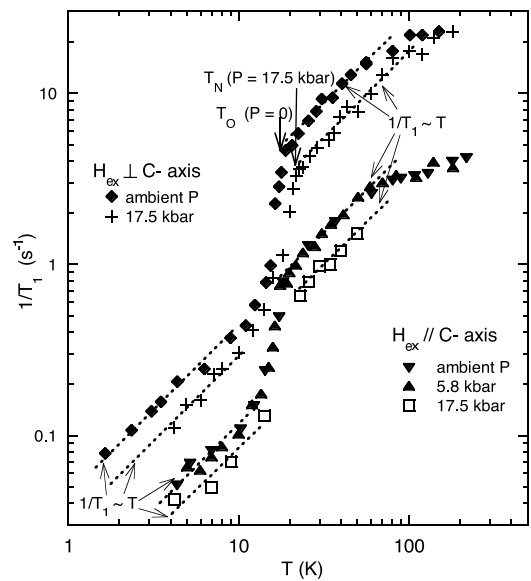}
\end{center}
\caption{
Temperature variation of the inverse of the NMR relaxation time $T_1$ at different pressure for $H \parallel c$. The nesting at $T_0$ or $T_{\rm N}$ is clearly visible. No drastic effect occurs at $T_{\rm x}$ (cited from ref.~\protect\citen{Koh86,Mat01} and drawn by Y. Kohori)
}
\label{fig:3}
\end{figure}

The case of PrFe$_4$P$_{12}$  is highly stimulating in the search to identify the OP of URu$_2$Si$_2$  at least of its low pressure phase. If U atom is in a tetravalent configuration U$^{4+}$, its configuration $5f^2$ will be quite similar to that of Pr$^{3+}$ with its $4f^2$ shell. Due to a strong hybridization, the local character of the U atoms is less preserved than that of Pr. Its Fermi Surface (FS) derived from band structure calculations has no obvious ``magic'' nesting wavevector as in the case of the skutterudite~\cite{Yam00,Ohk99,Har}; the number of carrier in the PM compensated regime has been estimated near 0.26  carrier (hole plus electron per U mole)  while, in the AF ground state,  $n$ is predicted to drop by a factor 3 in qualitative agreement with the NMR results of Fig.~\ref{fig:3}~\cite{Koh86}. There is no prediction on the FS change associated with an OP more exotic than the high pressure AF one with its $(1, 0, 0)$ wavevector. 

\begin{figure}[htb]
\begin{center}
\includegraphics[width=0.8 \hsize,clip]{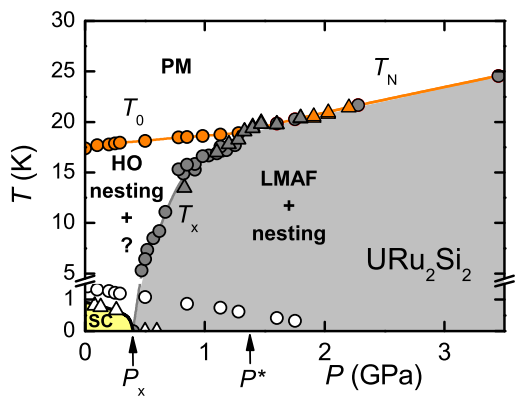}
\end{center}
\caption{
(Color online) High pressure phase diagram of URu$_2$Si$_2$ from resistivity
(circles) and ac calorimetry (triangles)~\protect\cite{Has}. 
The low pressure hidden order (HO) state is
characterized by a FS nesting which coexists probably with another order
parameter. At $P_{\rm x}$, the transition to the large moment antiferromagnetic (LMAF)
state is first order. Above $P^\ast$ only one transition is observed, however the nesting
character of the resistivity is preserved. Bulk superconductivity (SC) detected by ac
calorimetry (open triangles) is suppressed when the LMAF state appears. Open
circles present the temperature of the onset of the superconducting transition in the
electrical resistivity.
}
\label{fig:2}
\end{figure}

However, there are a lot of common points between the two systems. In the PM regime of URu$_2$Si$_2$, the local spin dynamic is smeared out in a large background but the crossing through the phase transition at $T_0$  leads to recover well defined excitations. A favourable mechanism for the slowdown of the spin dynamics is also the loss of carrier number. Below $T_0$, $n$ is at least below $0.1$~\cite{Koh86,Map86,Sch87,Bel04}. As for PrFe$_4$P$_{12}$, a $P$ tuning leads to switch in this case from HO to conventional AF~\cite{Ami07} (Fig.~\ref{fig:2}). However in URu$_2$Si$_2$, there is no indication either by NMR (Fig.~\ref{fig:3}) or by resistivity of a drastic change in the nesting conditions under $P$.~\cite{Mat01,Has} The persistence of nesting in URu$_2$Si$_2$  (whatever the ground state HO or AF ) is the required condition to reach slow spin dynamics which will favour multipolar phases with their corresponding sharp excitations. A previous $P$ invariance of nesting can be found in NMR experiments where the drop of $(T_1 T)^{-1}$ at $T_0$ visualizes clearly the drop in the carrier number as $(T_1 T)^{-1}$ varies like $\gamma^2$ and $\gamma$ as $n^{1/3}$ assuming a constant effective mass ($\gamma$ linear temperature term of the specific heat ). The results on both side of $P_{\rm x} = 0.5$ GPa are represented in Fig.~\ref{fig:3}~\cite{Mat01}.

Figure~\ref{fig:2} shows the (T, P) phase diagram of URu$_2$Si$_2$   determined by  ac calorimetry and resistivity measurements. The aim was to clarify if the ($T_{\rm x}$, $P_{\rm x}$) line which separates the HO ground state from AF ends up at a critical point $T_{\rm cr}$, $P_{\rm cr}$ or touches the ($T_0$, P) line which delimits the $P$ dependence of the $T_0$ transition from PM to HO. With our experimental accuracy, a tricritical point exist at $T^\ast$, $P^\ast$ where the three lines ($T_0$, $P$), ($T_{\rm N}$, $P$), ($T_{\rm x}$, $P$) converge. This implies that the HO phase has a different symmetry than the AF phase~\cite{Min05}.
Assuming that the time reversal symmetry is broken for both phases, one candidate for the HO is an octupolar OP~\cite{Agt94,Kis05} but up to now no corresponding feature has been reported. 

\begin{figure}[htb]
\begin{center}
\includegraphics[width=0.8 \hsize,clip]{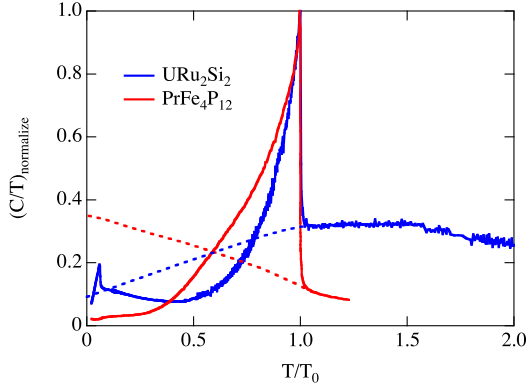}
\end{center}
\caption{
(Color online)
In the ratio $C/T$ of the specific heat by the temperature normalized to the value at the ordering temperature ($T_0$ or $T_{\rm A}$) as a function of reduced temperature for URu$_2$Si$_2$  and PrFe$_4$P$_{12}$ (after Ref.\protect\citen{YAok02}). The dashed lines show a rough extrapolation of $C/T$ in both cases if no phase transition occurs on the basis of entropy conservation.
}
\label{fig:4}
\end{figure}

An interesting point for further studies on PrFe$_4$P$_{12}$ is that in URu$_2$Si$_2$ below $T_0$ very sharp excitations 
($\Delta_{1, 0, 0}$ and $\Delta_{1.4, 0, 0}$) are observed at the wavevector $(1, 0, 0)$ and $(1.4, 0, 0)$. 
A full gap seems to be opened below $T_0$~\cite{Bro91,Bou93,Wie07} 
with respective energies $\Delta_{1, 0, 0} = 2\,{\rm meV}$ and  
 $\Delta_{1.4, 0, 0} = 4.5\,{\rm meV}$. 
On cooling, the intensity of the inelastic peak increases drastically just below $T_0$,
but then continue to rise gradually down to the superconducting temperature $T_{\rm c}$~\cite{Bou93}.
That leads to the strong decrease of the specific heat on cooling as shown in Fig.~\ref{fig:4}
which represents the renormalized $T$ dependence of $C/T$ 
in a reduced temperature scale $T/T_0$ and $T/T_{\rm A}$  for URu$_2$Si$_2$~\cite{Has,Agt94} and PrFe$_4$P$_{12}$~\cite{Mur86}. 
It is interesting to observe that for $T$ just below $T_0$, the initial drop of $C/T$ in URu$_2$Si$_2$  is faster than that for PrFe$_4$P$_{12}$. 
However, in PrFe$_4$P$_{12}$, $C/T$ continuously decreases on cooling.
In URu$_2$Si$_2$, $C/T$ goes through a minimum near $T_0 /3 \sim 7\,{\rm K}$
and starts to increase again at lower temperature;
the macroscopic low energy probe of the specific heat confirms
that the decoupling between exciton and quasi-elastic modes
is slow on approaching $T_{\rm c}$.
The increase of $C/T$ as $T$ decreases down to $T_{\rm c}$ is a characteristic of HFC
close to magnetic instability with their remarkable non Fermi liquid behaviour.

In the PM regime, the difference between a compensated (URu$_2$Si$_2$)~\cite{Kas07} and uncompensated (PrFe$_4$P$_{12}$) metal may play an important role. In the PM phase a simple counting of electrons predicts that PrFe$_4$P$_{12}$ is an uncompensated metal if the Pr atoms are in their trivalent configuration and that URu$_2$Si$_2$ is a compensated metal if  the U atoms are renormalized to their tetravalent configuration. In the game of localized or itinerant treatment of the 4\textit{f} electron, the Pr case presents the specificity that removing an even number of electrons from the 4\textit{f} shell will not change the condition of compensation while for Ce HFC, the change concerns only one carrier ($4f^1$); it leads to switch from uncompensated (LaRu$_2$Si$_2$) to compensated metal (CeRu$_2$Si$_2$ )~\cite{Onu96}. 
Within the hypothesis of trivalent Pr atoms, as in the ordered A phase the doubled unit cell contains an even number of electrons, PrFe$_4$P$_{12}$ should be also a compensated metal at low temperature. Experimentally, the situation is not clear. Deviations from the assumption that Pr is in a trivalent configuration imply also that the simple electron--hole symmetry is not valid.
As pointed out by Harima in this conference, PrFe$_4$P$_{12}$ is a challenging case for band structure calculations   
as $C_{\rm CF}$ is small and large mixing occurs. 
From the magnetoresistance behavior at low temperature it follows that URu$_2$Si$_2$ is a textbook illustration of a compensated semi-metal~\cite{Kas07} where even Shubnikov de Haas oscillations can be observed easily.
\begin{figure}[htb]
\begin{center}
\includegraphics[width=0.8 \hsize,clip]{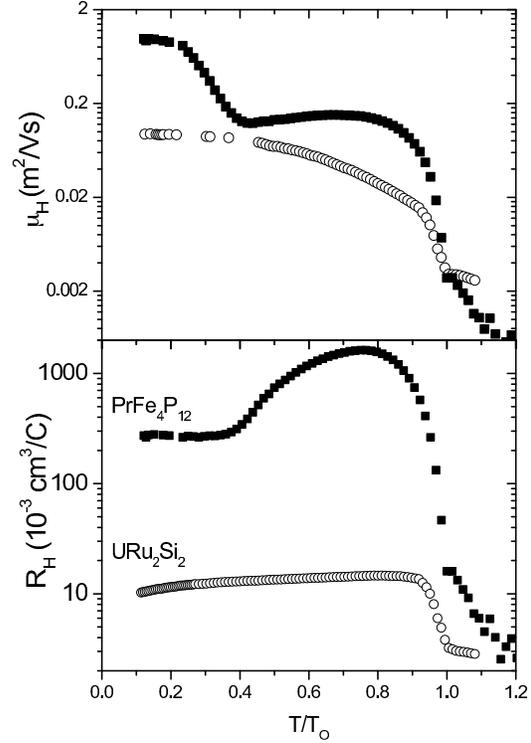}
\end{center}
\caption{
Mobility and Hall effect of PrFe$_4$P$_{12}$~\protect\cite{Pou06,Pou_proc} and URu$_2$Si$_2$~\protect\cite{Ohk99,Bel04} as a function of $T/T_0$. For PrFe$_4$P$_{12}$, there is a clear decoupling of the itinerant carrier with the exciton like excitations associated to the fancy multipolar ground state with a nmOP.
}
\label{fig:5}
\end{figure}

To our knowledge, no theoretical studies exist on the Kondo lattice taking into the singularities of a compensated and an uncompensated semi-metal. It is also worthwhile to remark that to preserve the entropy balance at $T_{\rm A}$ or $T_0$, for PrFe$_4$P$_{12}$, the extrapolation of the normal phase properties ($T_{\rm A}$ and $T_0$ collapsing ) will imply a heavy fermion state with strong non Fermi liquid behaviour, i.e. a strong increase of $C/T$ on cooling while for URu$_2$Si$_2$ it is required that $C/T$ passes through a broad maximum (see Fig.~\ref{fig:4}). Without HO the extrapolated PM regime of URu$_2$Si$_2$ will correspond to a weakly interacting system. 
  The difference in the behaviour of the remaining itinerant carrier between PrFe$_4$P$_{12}$ and URu$_2$Si$_2$ is also obvious in Fig.~\ref{fig:5} which represents the Hall mobility and the Hall constant as a function of reduced temperature~\cite{Pou06,Pou_proc,Sch87,Bel04}.  By comparison to URu$_2$Si$_2$, 
the number of carrier in PrFe$_4$P$_{12}$ has dropped by at least one order of magnitude at the phase transition.

In PrFe$_4$P$_{12}$, there is a clear low temperature decoupling  regime at $T_{\rm x} < T_{\rm A}$. For URu$_2$Si$_2$  there is a continuous  decrease of $R_{\rm H}$ at low temperature after the initial jump at $T_0$. The strong decoupling observed for PrFe$_4$P$_{12}$ is consistent with the observation of a nice $T^2$ resistivity term with a large $A$ coefficient
($A \sim 7\,\mu\Omega\,{\rm cm\, K}^{-2}$) already below $T_{\rm A}/2$~\cite{Tay07}.
The continuous shallow evolution of $C/T$ in URu$_2$Si$_2$  leads to the apparent observation of a $T^2$ term only below  $T_0 /10$ with $A \sim 0.1\,\mu\Omega\,{\rm cm\, K}^{-2}$.  A careful analysis shows that no simple law is observed above  the appearance of superconductivity at $T_{\rm c} \sim 1.2\,{\rm K}$~\cite{Has,Sch93,Mat_proc}. The fit of the low temperature resistivity with a power law $AT^\alpha$  gives an exponent $\alpha < 1.8$ close to $T_{\rm c}$. Furthermore $\alpha$ seems to decrease when the residual resistivity decreases~\cite{Sch93,Mat_proc}. Another paradox is that in URu$_2$Si$_2$ the relative weak number of carrier ($n < 0.1\,/{\rm U\, atom}$) does not seem to affect the Kadowaki--Woods ratio of $A/\gamma^2$ while for PrFe$_4$P$_{12}$, a large enhancement of $A$ is observed~\cite{Pou_proc} in good qualitative agreement with the prediction of ref.~\citen{Tsu05} for a  Kondo lattice assuming an unique spherical Fermi surface $A/\gamma^2 \sim n^{-4/3}$ or Fermi liquid arguments~\cite{Hus05} with $A/\gamma^2 \sim n^{-2}$.

The quasi insensitivity of the ratio of $A/\gamma^2$ in URu$_2$Si$_2$  suggests (1) either a bypass due to the compensated condition with for example light holes and heavy electrons~\cite{Kas07} having not only different $A$ terms but also different residual resistivities  (assumption usually rejected in HFC) or (2) a rather moderated nesting. The estimation of the carrier number is not straightforward and up to now only few parts of the Fermi surface have been detected (see ref.~\citen{Ohk99}). 

\begin{figure}[htb]
\begin{center}
\includegraphics[width=0.8 \hsize,clip]{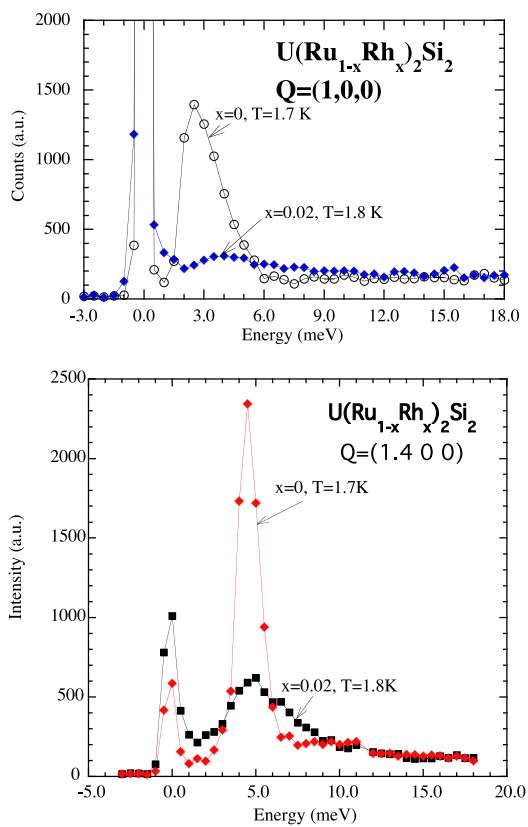}
\end{center}
\caption{(Color online) (a) Evolution of the inelastic neutron intensity~\protect\cite{Bou93} measured for $Q = (1, 0, 0)$ and (b) for  $Q = (1.4, 0, 0)$ for pure URu$_2$Si$_2$  case and weakly doped U(Ru$_{0.98}$Rh$_{0.02}$)$_2$Si$_2$ case at $T \sim 1.7\,{\rm K}$.
}
\label{fig:6}
\end{figure}

Another common point with PrFe$_4$P$_{12}$ is the high sensitivity of URu$_2$Si$_2$ to doping. 
The sharp excitations $\Delta_{1, 0, 0}$ and $\Delta_{1.4, 0, 0}$ 
are strongly smeared out with weak Rh doping on Ru side~\cite{Bou93}. 
The sharp inelastic feature at $\Delta_{1, 0, 0}$ is strongly damped below $T_0$
even at $T=8.1\,{\rm K}$ in the HO phase just on entering in LMAF phase at $T_{\rm N}=8\,{\rm K}$.
The weak doping with $2\,{\%}$ leads to end up in an LMAF ground state
quite similar to the one observed under high pressure above $P_{\rm x}$
Figure~\ref{fig:6} shows at $1.7\,{\rm K}$ the drastic wipe out for $x = 2\,{\%}$
of the inelastic neutron intensity.
Very recently the collapse of the intensity of the low energy excitations at $Q = (1, 0, 0)$
was directly demonstrated under a pressure of $0.7\,{\rm GPa}$
in the pure compound URu$_2$Si$_2$ on entering in the LMAF phase~\cite{Vil08}.
Focusing on the high sensitivity to HO to doping,
in agreement with the strong damping of the excitations with doping,
the specific heat anomaly at $T_0$ is strongly reduced~\cite{Bou93,Bou05}.
There is also a corresponding large change in the thermal conductivity bump observed at $T_0$~\cite{Sha06}.
The necessity to achieve a clean limit for the HO phase of URu$_2$Si$_2$  
and the nmOP phase of PrFe$_4$P$_{12}$  is quite reminiscent 
of the examples of unconventional superconductivity 
where any impurity will be efficient to break the phase coherence 
due to the change of sign of the OP.

\begin{figure}[htb]
\begin{center}
\includegraphics[width=0.8 \hsize,clip]{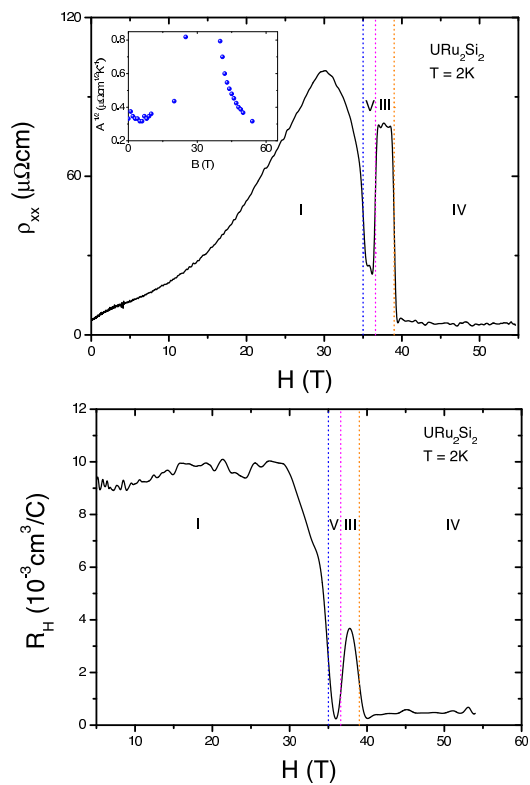}
\end{center}
\caption{
(Color online)
(a): Variation of the transverse resistivity $\rho_{xx}$~\protect\cite{Leva} as a function of the magnetic field $H$ up to $55\,{\rm T}$ at $2\,{\rm K}$. The insert shows the field variation of the $A^{1/2}$ term of the $T^2$ dependence of the resistivity.
(b): Field dependence of the Hall constant~\protect\cite{Leva}. The label of ordered phase I, V, III is from reference~\protect\cite{Kim03}.
}
\label{fig:7}
\end{figure}

In both cases, increasing the magnetic polarization under magnetic field leads to a polarized paramagnetic metal (PPM) with a large number of carriers. For PrFe$_4$P$_{12}$, $(1, 1, 1)$ is a singular direction as for this direction a CF level crossing occurs~\cite{Tay04}. For URu$_2$Si$_2$  the situation is more complex. For a field $H \parallel c$ applied in the easy magnetization axis, a cascade of ordered ground states exists referred as I (HO phase) III, V (still unidentified ordered phases) and IV the PPM limit according to the ref.~\citen{Kim03}. Using the pulsed high magnetic field facility in Toulouse, new sets of results on transverse magnetoresistivity $\rho_{\rm xx}$, Hall and Nernst effects have been obtained~\cite{Leva}. At first glance they confirm previous data~\cite{YOh07,YJo07} however the high accuracy helps to demonstrate that the crossing I $\to$ V $\to$ III $\to$ IV with $H$ have clear signature in the three quantities notably as reproduced Figs.~\ref{fig:7}(a) and \ref{fig:7}(b) in $\rho_{xx}$ and $R_{\rm H}$. The insert of Fig.~\ref{fig:7}(a) shows the field variation of the $A$ coefficient. Its derivation above $H=15\,{\rm T}$ in the HO regime is ambiguous as already below $6\,{\rm K}$ in this compensated semi-metal, the temperature dependence of $\rho$ is dominated by the quantum orbit motion and not by the collision between quasiparticles. One goal will be to try to go from the field dependence of $A$ to the field dependence of average effective mass using a relation between $A$, $\gamma$ and $n (H, P)$. However, we have already mentioned that even at $H = 0$ a simple formula with a single spherical Fermi surface does not held.

From Hall effect measurements, the estimation in the carrier drop below $T_0$
but also from the previous band calculation in the PM regime just above $T_0$
the number of carrier per U atom is near $0.3$.
Above $H=H_{\rm M}$, the high magnetic field Hall effect measurement 
(comparison of Fig.~\ref{fig:5} and \ref{fig:7}(b))
indicates an increase of $n$ from the PM to PPM phase;
in the PPM state $n$ is nearly unity.
Concerning the $H$ change in interaction, the situation of URu$_2$Si$_2$ is
rather similar to that observed in CeRu$_2$Si$_2$~\cite{Flo05}. 
For PrFe$_4$P$_{12}$ as the field variation of the number of charge carriers 
will be at least two orders of magnitude at $H_{\rm A}$, 
the main change occurs just at $H_{\rm A}$. In the PPM, the $\gamma$ coefficient decreases smoothly with $H$. This indicates mainly that PrFe$_4$P$_{12}$ is a Kondo system with a rather high Kondo temperature almost like CeSn$_3$ and CeBe$_{13}$ for Ce HFC~\cite{Loe93,Flo05}. The $H$ restoration of a large number of carriers leads to recover an intermediate valence regime detected above $T_{\rm A}$ at $H=0$; a similar behaviour has been recently observed in CeRh$_2$Si$_2$ just above its metamagnetic transition at $H_{\rm M} \sim 25\,{\rm T}$~\cite{Levy}.

An additional interest in URu$_2$Si$_2$  is the appearance of superconductivity (SC)~\cite{Sch86}. Recent ac calorimetry~\cite{Has} results confirm that bulk superconductivity does not coexist inside the AF phase~\cite{Ami07,Sat06}. 
As it has been proposed for UPd$_2$Al$_3$~\cite{Miy01,Sat01,Cha07}, 
the appealing possibility is that SC comes from the excitonic mode at $\Delta_{1, 0, 0} \sim 2\,{\rm meV}$ 
which was suggested to collapse in AF phase~\cite{Has}. 
A new set of inelastic neutron measurements under pressure has established this point just recently.~\cite{Vil08} 
In resistivity measurements, SC seems to coexist with AF up to $2\,{\rm GPa}$ in the present experiment. There is no understanding on the origin of this residual superconductivity as well as there is also no clear view why inside the HO phase, a tiny sublattice magnetization related to surviving AF droplets is generally associated with the HO transition at $T_0$. This rises the question of the dominant lattice defects.
The doubt still persists of an intrinsic residual tiny sublattice magnetization in the HO phase. 
Despite two decades of studies on URu$_2$Si$_2$, 
the criteria for the sample purity remain ambiguous. 
One experimental paradox in the recent published data 
on high quality single crystal (${\rm RRR} =600$)~\cite{Kas07} 
is that the SC specific heat and resistivity anomalies are rather broad~\cite{Mat_proc}. 
A further evidence of residual imperfections is that a large broadening still persists in the resistive SC transition close to the superconducting upper critical field $H_{\rm c2}(0) \sim 3\,{\rm T}$ for $H \parallel c$ (comparison of ref.~\citen{Kas07} with ref.~\citen{Sch93}). 
These broadenings may be generated by intrinsic effects as stacking faults.

\section{SmB$_6$: valence, metal-insulator transition and magnetism}
Progress on high pressure and also on complementary microscopic measurements on synchrotron facilities has clarified the interplay between the valence mixing, Kondo effect, metal-insulator transition and magnetic long range order (LRO) in the two well-known cases of SmS~\cite{Der06,Imu06} and SmB$_6$~\cite{Der06,Der}. The simple image is that the valence mixing between the two Sm$^{2+}$ and Sm$^{3+}$ configuration plays a key role according to the relation:
\begin{equation}
{\rm Sm}^{2+} \longleftrightarrow {\rm Sm}^{3+} + 5d.
\end{equation}

In the extreme conditions of Sm$^{2+}$ or Sm$^{3+}$ the only selected configurations of the ground states are respectively a PM insulator and a LRO metal; the isostructural trivalent lanthanum references LaS and LaB$_6$ are excellent metals with $n \sim 1$. The paradox is that despite the valence ($v$) is intermediate as estimated by high energy spectroscopy~\cite{Ann06,Dal05} or Raman measurements~\cite{Ogi05}, the development of the correlation on cooling will lead to renormalize the electronic and magnetic properties either to the divalent limit (insulator ground state with 2+ magnetic form factor) for $v < 2.8$ or to the trivalent limit for $v > 2.8$ (metallic ground state with 3+ magnetic form factor)~\cite{Der06}.  The same renormalization has been observed for TmSe. Just before we have discussed the apparent renormalization of URu$_2$Si$_2$ to the U$^{4+}$ configuration.

\begin{figure}[htb]
\begin{center}
\includegraphics[width=0.8 \hsize,clip]{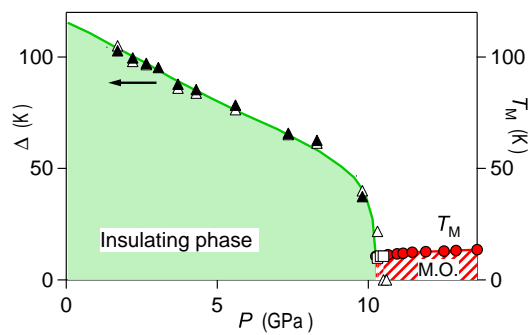}
\end{center}
\caption{
(Color online)
$(T, P)$ phase diagram of SmB$_6$~\protect\cite{Der06,Der}. The charge gap $\Delta$ is derived from the resistivity and the onset of LRO magnetic detected by specific heat and resistivity measurements. The magnetic ordered state coincides with the insulator metal transition.
}
\label{fig:8}
\end{figure}

For illustration, the $(T, P)$ phase diagram of SmB$_6$ is shown in Fig.~\ref{fig:8}. The decrease of the charge gap has been derived from resistivity data. The $P$ variation of LRO temperature $T_{\rm N}$  has been obtained from ac calorimetry and resistivity measurements. Microscopic confirmation of homogeneous LRO has been provided by nuclear forward scattering~\cite{Bar05}. At $P_{\Delta}$, $v$ is estimated to be close to $2.8$. In contrast to the case of Ce where the hybridization is large due to a larger expansion of its $4f$ shell than for that of Sm, in Sm HFC
moderated values of $T_{\rm K}$  can be achieved even for rather strong difference from $n_f \sim 1$ for the Sm$^{3+}$ configuration as $T_{\rm K}$: 
\begin{equation}
T_{\rm K} \sim  (1 - n_f )\Delta
\end{equation}
(As $n_f = 1$ corresponds to the pure trivalent configuration
(respectively 1, 6 and 13 electrons in the $4f$ shell of Ce, Sm and Yb),
for example in Sm HFC, $n_f$ is related to the valence $v$ via the relation $v = 2 + n_f$; $n_f = 0$
for a pure divalent configuration as it occurs for the black phase of SmS.
In the case of Ce HFC, $v = 4 - n_f$ and for Yb HFC like in Sm HFC $v = 2 + n_f$).
 
The  width $\Delta$ of the $4f$ virtual bound state is related to the square of the hybridization potential. Basically for $n_f \sim 0.8$ in SmB$_6$  as in SmS,  $k_{\rm B}T_{\rm K}$ becomes smaller than the CF energy and thus magnetic ordering appears.

\section{Comparison with different rare earth skutterudites: SmOs$_4$Sb$_{12}$, SmFe$_4$P$_{12}$, CeOs$_4$Sb$_{12}$, CeFe$_4$P$_{12}$}  
\begin{figure}[htb]
\begin{center}
\includegraphics[width=0.8 \hsize,clip]{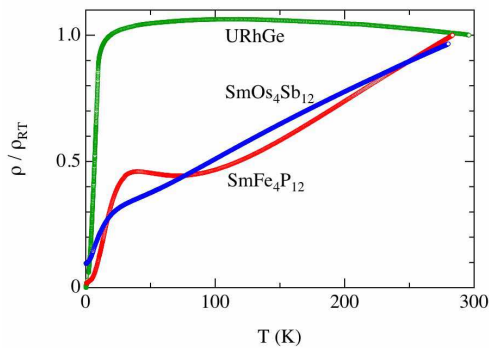}
\end{center}
\caption{
(Color online)
T dependence of the resistivity of the two FM anomalous Sm skutterudite SmOs$_4$Sb$_{12}$ ($v = 2.8$) and SmFe$_4$Sb$_{12}$ ($v = 3$) drawn by H. Sugawara by comparison to URhGe (D. Aoki).
}
\label{fig:9}
\end{figure}

Figure~\ref{fig:9} shows the resistivity as a function of $T$ for two metallic anomalous FM skutterudites SmOs$_4$Sb$_{12}$  and SmFe$_4$P$_{12}$ in comparison to URhGe an exotic FM/HFC superconductor~\cite{Flo05}. Recent experiments on SmOs$_4$Sb$_{12}$ by soft and hard X ray spectroscopy~\cite{Yam07} and X ray absorption spectroscopy~\cite{Miz07} show clearly that the situation of SmOs$_4$Sb$_{12}$ is quite comparable to that of SmB$_6$. Even for $v \sim 2.8$, FM occurs at low temperature with huge residual $\gamma$ term $\sim 800\,{\rm mJ\,mole^{-1}K^{-2}}$ at $T_{\rm Curie} = 3\,{\rm K}$~\cite{Kot05} and a sublattice magnetization $M_{\rm o}$ near $0.04\,\mu_{\rm B}$. Under pressure, the FM resistivity anomaly becomes visible and $T_{\rm Curie}$ increases continuously reaching $10\,{\rm K}$ at $P=4\,{\rm GPa}$~\cite{Sam05}. At $P = 0$, SmOs$_4$Sb$_{12}$ is very close a FM instability. 

\begin{figure}[htb]
\begin{center}
\includegraphics[width=0.8 \hsize,clip]{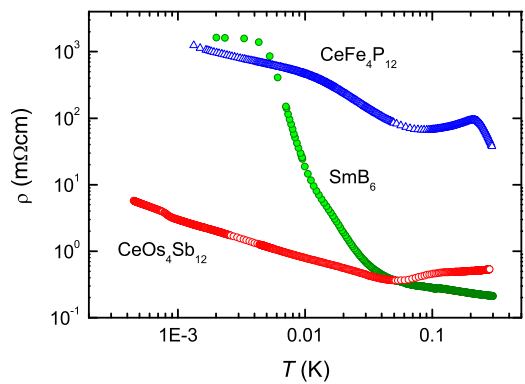}
\end{center}
\caption{
(Color online)
$T$ dependence of resistivity of the two Ce Kondo insulator skutterudite CeOs$_4$Sb$_{12}$ and CeFe$_4$Sb$_{12}$. The first one is an AF and the second one a PM state. For comparison we have drawn the result of SmB$_6$ at $P=0$.
}
\label{fig:10}
\end{figure}

Decreasing the hybridization leads to the interesting case of FM of SmFe$_4$P$_{12}$~\cite{Tak03} 
where $n_f \sim 1$ and $\gamma = 300\,{\rm mJ\,mole^{-1}K^{-2}}$ and $M_{\rm o}= 0.17\,\mu_{\rm B}$.
Under pressure, the FM of SmFe$_4$P$_{12}$  is robust at least up to $8\,{\rm GPa}$ while the $A$ coefficient decreases only by a factor 2 from $0.24\,\mu\Omega{\rm cm\,K^{-2}}$ at $P=0$ to $0.14\,\mu\Omega{\rm cm\,K^{-2}}$ at $P=8\,{\rm GPa}$~\cite{Kur07}. For these two systems, dominant FM interactions preclude an insulating ground state as it is the case for TmTe in its IV phase ($v\sim 2.6$)~\cite{AYam07} by contrast to TmSe where AF is associated with a metal insulator transition.
Figure~\ref{fig:10} shows the resistivity of two skutterudite Kondo insulators  CeOs$_4$Sb$_{12}$ and
CeFe$_4$P$_{12}$ incomparison to SmB$_6$ at $P=0$. The first one is a fancy case known to end in a semi conductor ground state but on the AF side just close to an AF quantum critical point as its low N\'{e}el temperature ($T_{\rm N} \sim 0.8\,{\rm K}$) indicates~\cite{Yog05,Iwa07}. Pressure measurements~\cite{Hed03} show that the insulating tendency increases with $P$ but no studies has been reported on the pressure dependence of $T_{\rm N}$. In agreement with the observation that the shrinking of the lattice parameter drives to an insulating ground state CeFe$_4$P$_{12}$ appears as a robust insulator~\cite{Sug06}. However quite surprisingly a weak substitution of Ce by few percent of La leads to a resistivity drop by one order of magnitude. 

\section{Large domain for the future studies}
Many other cases of anomalous rare earth compounds can be examined. For each of them, a large variety of experiments need to be performed. Let us mention a few proposals.

At least, one of the priorities is to increase the knowledge on the beautiful case of PrFe$_4$P$_{12}$. One important goal is to get accurate inelastic neutron scattering spectra to identify the specific excitations of the nmOP as well as their temperature variations. Another key point is to play with the remaining itinerant electrons. Why do they not enjoy forming Cooper pairs? Furthermore an interesting aspect is that the exciton mode of the multipolar OP may play a main  role in thermal transport. How to distinguish from the classical phonon contribution? The proof of a compensation or not below $T_{\rm A}$ must be given as it has consequences on the correct treatment of the Kondo effect in Pr HFC~\cite{Yot02,Ots05}

On the other examples, there is a need of basic simple experiments such as thermal expansion in SmOs$_4$Sb$_{12}$ for a comparison with the valence drop in temperature~\cite{Yam07}, in CeOs$_4$Sb$_{12}$ for a basic test of the proximity to AF QCP. Why FM systems like SmOs$_4$Sb$_{12}$, SmFe$_4$P$_{12}$ cannot become superconductors like UGe$_2$ or URhGe?~\cite{Flo05}  Behind these challenges, there is the underlining question of the required purity as
a high sensitivity to doping has been observed for many cases.
Few or almost no deep studies concern the filling factor despite the fact that in a large cage system like PrOs$_4$Sb$_{12}$ or SmOs$_4$Sb$_{12}$ there may be a real problem. It is now time to clarify the situation. In the case of PrFe$_4$P$_{12}$, fortunately the filling factor is believed to be very near 1. However the search for new effects will certainly require material improvements.

\section*{Acknowledgments}
J. Flouquet likes to thank H. Harima and H. Sato for entering in the skutterudite ``club'' and continuous stimulating explanations. The work on URu$_2$Si$_2$ starts also by the simulation of my friend's works, K. Asayama and Y. Miyako. K. Miyake was always very available and open to clarify basic questions. Discussions with Y. Haga were very useful to precise the material reality notably on URu$_2$Si$_2$. We thanks H. Hidaka, Y. Kohori and H. Sugawara for the drawing of Figs.~\ref{fig:1}, \ref{fig:3}, \ref{fig:9} and \ref{fig:10}. We appreciate all discussions with many participants


\end{document}